# Probing the nature of the conjectured low-spin wobbling bands in atomic nuclei

S. Guo[1,2], X.H. Zhou[1,2]✉, C. M. Petrache[3], E. A. Lawrie,[4, 5] S. H. Mthembu,[4, 5] Y. D. Fang[1, 2],

H. Y. Wu[6], H. L. Wang[7], H. Y. Meng[7], G. S. Li[1, 2], Y. H. Qiang[1], J. G. Wang[1, 2], M. L. Liu[1, 2],

Y. Zheng[1, 2], B. Ding[1, 2], W. Q. Zhang[1, 2], A. Rohilla[1], K. R. Mukhi[1], Y. Y. Yang[1, 2], H. J. Ong[1, 2],

J. B. Ma[1, 2], S. W. Xu[1, 2], Z. Bai[1, 2], H. L. Fan[8], J. F. Huang[9], J. H. Li[1], J. H. Xu[1], B.F. Lv[1],

W. Hua[10], Z. G. Gan[1, 2], and Y. H. Zhang[1,2]

**Abstract**

**Precession is a unique motion in which the orientation of the rotational axis of a rotating body is not fixed but moving, and it generally exists in the Universe from giant stars through tiny atomic nuclei. In principle, the precession of an atomic nuclide can be approximately described as wobbling motion, arising from the coupling of a rotation and a harmonic vibration. Recently, a number of wobbling bands were reported at low spin, which violate the wobbling approximation that can be valid only at high spin. Here we explore the nature of the reported low-spin wobbling bands. Via a new experiment, we demonstrate that one such band in $^{187}$Au is generated by dominant single-particle excitation rather than by the excitation of a wobbling phonon. We point out that the imperfect research paradigm used previously would lead to unreliable identification of low-spin wobbling bands. Consequently, new experimental approaches should be developed to distinguish among the different excitation mechanisms that can give rise to the observed low-spin bands in odd-even nuclei.**

Introduction

Precession is well known as a motion of a macroscopic rigid body, where its rotational axis describes a cone around a fixed direction. Actually, it appears in quite different scenes such as atomic nuclei of many-body quantum microscopic systems, for which the external force can be neglected and the moments of inertia are between hydrodynamic and rigid types. Nuclear precession was predicted by Davydov and Filippov[1], and described as nuclear wobbling in an approximate way by Bohr and Mottelsson[2]. When a triaxially deformed nucleus precesses, one of the three principal nuclear axes plays the role of the axle, and rotates around the space-fixed angular momentum. In the body-fixed frame of a nucleus, one can observe the precession of the angular momentum. As shown in Figure

[1]*Key Laboratory of High Precision Nuclear Spectroscopy, Institute of Modern Physics, Chinese Academy of Sciences, Lanzhou 730000, People's Republic of China*
[2]*School of Nuclear Science and Technology, University of Chinese Academy of Sciences, Beijing 100049, People's Republic of China*
[3] *Université Paris-Saclay, CNRS/IN2P3, IJCLab, 91405 Orsay, France*
[4]*iThemba LABS, National Research Foundation, PO Box 722, 7131 Somerset West, South Africa*
[5]*Department of Physics & Astronomy, University of the Western Cape, P/B X17, Bellville ZA-7535, South Africa*
[6]*State Key Laboratory of Nuclear Physics and Technology, School of Physics, Peking University, Beijing 100871, China*
[7]*School of Physics and Microelectronics, Zhengzhou University, Zhengzhou 450001, China*
[8]*College of Engineering Physics, Shenzhen Technology University, Shenzhen 518118, China*
[9]*Guangzhou Municipal Ecological Environment Bureau Huadu District Branch, Guangzhou 510800, People's Republic of China*
[10]*Sino-French Institute of Nuclear Engineering and Technology, Sun Yat-Sen University, Zhuhai 519082, China*
✉*e-mail: zxh@impcas.ac.cn*

1, nuclear precession occurs when the rotational angular momentum $R'$ is not oriented along a principal nuclear axis (shown with dashed line), but is tilted away and precesses. The total angular momentum $J'$ resulting from the coupling of $R'$ and the single-particle angular momentum $j$ also precesses around the nuclear axis. Within the wobbling approximation, wobbling phonons are introduced to describe a quantized small-angle deviation of the axle with respect to the direction of $J'$. Such approximation is only valid at high spin, where it can be regarded as a harmonic vibration. At low spin, the deviation angle is too large to satisfy the wobbling approximation, and therefore the description of the rotational bands in terms of wobbling phonon excitations is invalid.

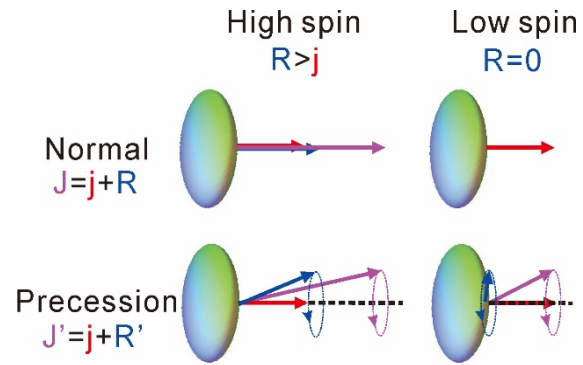

**Fig. 1: Angular momentum geometry of nuclear precession.** Normal (upper) and precession (lower) modes for high-spin (left) and low-spin (right) systems are sketched in the body fixed frame of an odd-A triaxial nucleus.

Nuclear wobbling was first reported in the 160-mass region, where all identified wobbling bands were observed at high spin[3-8]. However, a number of rotational bands were interpreted recently as wobbling excitations at low spin[9-15]. These bands were described as transverse or longitudinal wobbling by Frauendorf and Dönau[16], which is achieved when the angular momentum of the odd quasiparticle is rigidly aligned with one of the principal axes of the rotating triaxial core. The spins of the lowest observed states in most of those bands are $J=j+1$, where $j$ is the angular momentum of the single quasi-particle (see Fig. 1). This means that in such low-spin states the wobbling motion can occur for an extremely slow collective rotation ($R'$). However, this is in contradiction with the approximation condition for wobbling motion, which requires large angular momentum, as pointed out by Bohr and Mottelsson for even-even nuclei and by Frauendorf and Dönau for odd-mass nuclei, and recently demonstrated in the work of Lawrie et al.[17].

It is therefore necessary to explore the true excitation mechanisms generating the low-spin bands interpreted as wobbling excitations. The present new experimental results on the previously proposed wobbling band in $^{187}$Au, unequivocally show that the band is generated by dominant single-particle excitation and cannot be associated with wobbling. Furthermore, from an overall assessment, we conclude that the experimental and theoretical evidence for the reported low-spin wobbling bands is generally insufficient. To understand the mechanisms giving rise to those bands, convincing experimental and theoretical examinations are required.

**Results**

*Re-investigation on the reported low-spin wobbling band in $^{187}$Au*

In order to unveil the nature of the low-spin bands in $^{187}$Au, an experiment was carried out at the Heavy Ion Research Facility in Lanzhou (HIRFL)[18,19], China. The excited states of $^{187}$Au were populated using the $^{175}$Lu ($^{18}$O, 6n) reaction with a 108-MeV $^{18}$O beam and a stack of two self-supported natural Lu targets with a thickness of 0.7 mg/cm$^2$ each. The $\gamma$ rays were detected with a detector array consisting of 8 segmented clover detectors and 16 coaxial High-Purity Germanium (HPGe) detectors. Eight of the HPGe detectors were equipped with anti-Compton shields. All clover detectors were placed in a ring perpendicular to the beam direction. The 16 HPGe detectors were placed in four rings at 26°, 52°, 128° and 154° with respect to the beam direction. Approximately, $5\times10^{10}$ double- or higher-fold events were recorded using a general-purpose digital data acquisition system[20].

Fig. 2a shows a partial level scheme of $^{187}$Au built in the present work, together with a partial level scheme of $^{188}$Pt. Bands 1-4 established in previous work[21,22] are confirmed. The structure in pink was reported in Ref. [14], and interpreted as the unfavored signature branch of the $\pi h_{9/2}$ band (Band 1) in $^{187}$Au. It is worth noting that the 265-, 405-, 412-, and 549-keV transitions have very similar energies to those of the 265-, 405-, 414-, 551-keV transitions of $^{188}$Pt[23] (the sequence in green in Fig. 2a), which is also populated in the reaction. Our data shows no sign for the existence of the previously proposed 429- and 437-keV linking transitions, as shown in Fig. 2b. In addition, this structure was not identified in the detailed β-decay study of $^{187}$Hg performed by Rupnik et al.[24], in which plenty of excited states in $^{187}$Au with spin up to I = 19/2 were observed. If this band did belong to $^{187}$Au and did have the features reported in Ref [14] in terms of quasiparticle configuration, spin, parity, and very low excitation energy, it would be present in our data and would have been observed by Rupnik et al.[24].

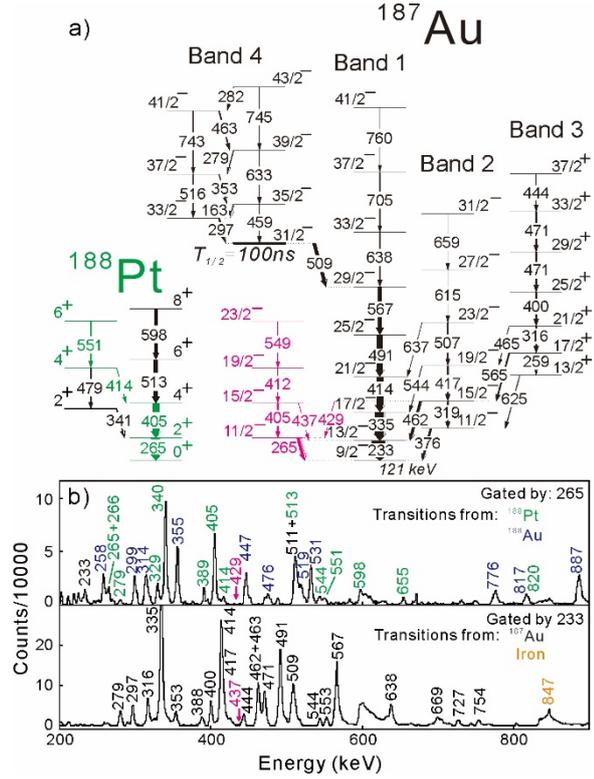

**Fig. 2 Level scheme and spectra. a** Partial level scheme of $^{187}$Au and $^{188}$Pt. The structure in pink was reported in Ref. [14], but not observed in the present data. Instead, a similar sequence in green was assigned to $^{188}$Pt. **b** Spectra gated by the 265- and 233-keV transitions.

Band 2 was first proposed as the unfavored signature branch of Band 1 built on a single-particle excitation[21], but interpreted later as a collective wobbling band[14]. Band 2 consists of a series of excited states with spin increasing in steps of 2, and decays to Band 1 via several $\Delta I = 1$ transitions. Usually, $\Delta I = 1$ transitions are mixed, with mainly electric quadrupole (*E2*) and magnetic dipole (*M1*) components, in proportion described by the *E2/M1* mixing ratio ($\delta$). If Band 2 is a wobbling band, the $\Delta I = 1$ transitions are expected to be dominated by the *E2* components, giving rise to larger than one absolute values of $\delta$. This is contrary to the expectation of *M1* dominated character of the transitions between signature partner branches.

In this work, the $\delta$ values of the transitions linking Band 2 to Band 1 were determined using two complementary measurements of a two-point angular correlation ratio, $R_{ac}$, and of linear

polarization, P. $R_{ac}$ was deduced from the ratio of the γ-ray intensities in the spectra measured by the 154° and 90° detectors, gated on transitions observed in all detectors. P was deduced from the Compton scattering events in perpendicular and horizontal directions in adjacent crystals of the clover detectors (see Methods section for the procedures). The measured polarization of the 376-keV, $11/2^- \rightarrow 9/2^-$ transition is in good agreement with the value of -0.10(5) reported previously[21]. The δ values were deduced from the measured values of $R_{ac}$ and P, as shown in Fig. 3. The curves are sensitive to the σ/I parameter in which σ is the variance of the projection of the angular momentum I along the beam direction. The σ/I parameter might be affected by states with lifetimes of the order of a picosecond or larger. In the present work, δ values are deduced with the σ/I parameter not restricted. This induces a larger uncertainty in the measured δ, however the result does not depend on an assumed value for σ/I. Despite the large error bars, it is clear that the absolute values of the deduced δ values are smaller than 1, indicating dominant *M1* nature of these transitions. Data points that correspond to the large δ values deduced in Ref. [14] are shown in red in Fig. 3. They are in distinct disagreement with both the presently deduced values and the values obtained from previous internal conversion coefficient measurements[24,25]. Considering the measured $B(E2)_{in}$ values of 100 - 200 W.u. in the collective bands of $^{187}$Au and the $^{186}$Pt core[27,28], the values of $B(E2)_{out}$ are estimated to be only a few W.u., which is inconsistent with the proposed wobbling interpretation for Band 2[14], and even with any interpretation that involves a collective type of excitation of Band 2 with respect to Band 1.

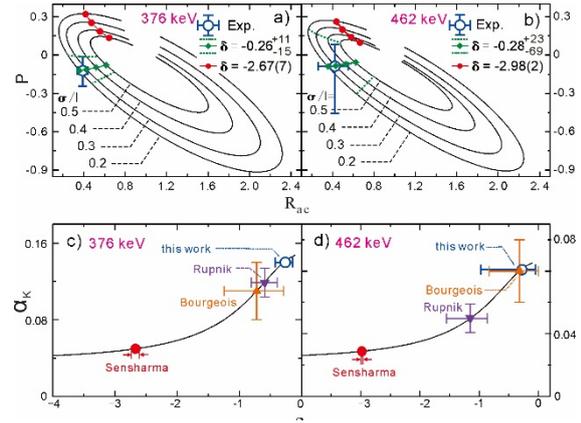

**Fig. 3 Experimental mixing ratios. a and b** Plots of polarization P versus $R_{ac}$ for different values of δ and σ/I, together with experimental results and estimated results for certain δ values. **c and d** Internal conversion coefficients $α_K$ as a function of δ, together with the experimental results from the present work and those from Sensharma et al.[14], Rupnik et al.[24], and Bourgeois et al.[25]. The δ values for the Refs. [24] and [25] were deduced from the reported internal conversion values using the BrIcc code[26].

In order to study further the nature of Band 2, one-quasiparticle-plus-triaxial-rotor (QTR) model calculations using the code of P. Semmes and I. Ragnarsson[29] were carried out (see the details in Methods section). Standard parameters are used for the Nilsson potential[30] and for the pairing strength. The γ-dependence of the moments of inertia of the core has irrotational-flow nature, as suggested by an empirical evaluation[31]. The spin dependence is taken into account assuming variable moments of inertia described by the Harris parameters $J_0 = 25$ $\hbar^2$MeV$^{-1}$ and $J_1 = 8$ $\hbar^4$MeV$^{-3}$. The adopted quadrupole deformation of $\varepsilon_2=0.21$ is similar to that previously used in Refs. [14, 23]. The adopted triaxial deformation is $\gamma = 12°$, as suggested in the theoretical calculations for the $^{186}$Pt core[32]. The calculated proton Fermi level is near the highest $h_{11/2}$ and lowest $h_{9/2}$ orbitals, as in Ref. [21]. Nine negative-parity orbitals adjacent to the Fermi level are included in the calculations.

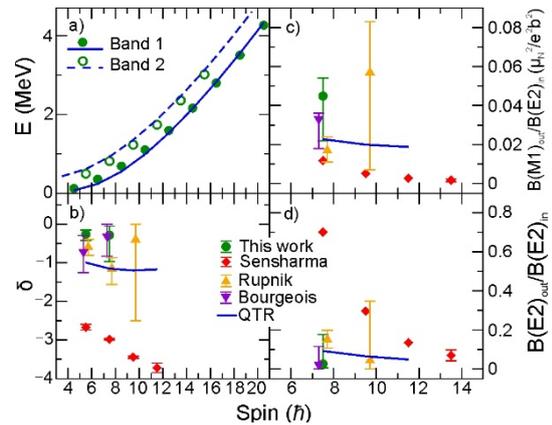

**Fig. 4 Comparison between experimental and calculated results. a** Excitation energies (E), **b** mixing ratios (δ), **c** $B(M1)_{out}/B(E2)_{in}$ ratios, and **d** $B(E2)_{out}/B(E2)_{in}$ ratios. The values associated with the previously measured δ values [14, 21, 24, 25] are also shown. Some points are slightly shifted horizontally to avoid overlapping.

The calculated excitation energies of the two lowest-energy rotational bands with $h_{9/2}$ nature are shown in Fig. 4a. The wave functions clearly indicate that Bands 1 and 2 are based primarily on the lowest and second-lowest energy orbitals originating from the $h_{9/2}$ subshell, respectively, suggesting that the excitation between the two bands has a single particle character. The calculated $\delta$ values, $B(M1)_{out}/B(E2)_{in}$ and $B(E2)_{out}/B(E2)_{in}$ ratios are in good agreement with our data and with the data from Refs. [20, 23, 24], but in contrast with those from Ref. [14] (see Figs. 4b, 4c and 4d). This agreement further supports the proposed non-collective character of the excitation, for more details, see the Methods section.

*Assessment on the reported low-spin wobbling bands*

In order to check if the risk of misinterpretation also exists in other reported low-spin wobbling bands, we assessed the published experimental and theoretical evidence. The critical experimental proof in favor of the wobbling interpretation is the observation of predominant *E2* character for the *ΔI=1* transitions linking the proposed one-phonon wobbling bands to the corresponding zero-phonon bands, which is determined by measuring the *δ(E2/M1)* mixing ratios. Additional support for wobbling motion is provided by a comparison between calculated and experimental results on excitation energies and ratios of transition probabilities.

The $\delta$ values can be deduced from angular distribution and/or linear polarization measurements. Angular distribution curves are calculated assuming appropriate values for $\delta$ and $\sigma/I$. The suitable $\delta$ values, which correspond to good agreement between the calculated curves and the experimental points, are obtained. As shown in Fig. 5a, the angular distribution curves become steeper and steeper with decreasing $\delta$ value below 0. However, for $\delta < -1$ the slope decreases. There are usually two minima for the $\chi^2$ values corresponding to two $\delta$ solutions $|\delta_1|>1$ and $|\delta_2|<1$ (see Fig. 5b). In some cases, almost equally good fits are obtained for both solutions (see Fig. 5c), leading to similar $\chi^2$ minima. Sometimes, one solution can fit the experimental results better than the other one (see Fig. 5d). It is therefore important to discuss the confidence level of $\chi^2$, and check whether the second solution for $\delta$ can be confidently excluded based on the $\chi^2$ fit. It is worth noting that the number of counts in the peaks used to obtain the angular distribution are not measured directly, but deduced from the γ-γ coincidence matrices. Both the central values and the uncertainties can be affected by the selection of parameters in the process of creating and analyzing the gated spectra. The resulting uncertainties of $\chi^2$ should be considered.

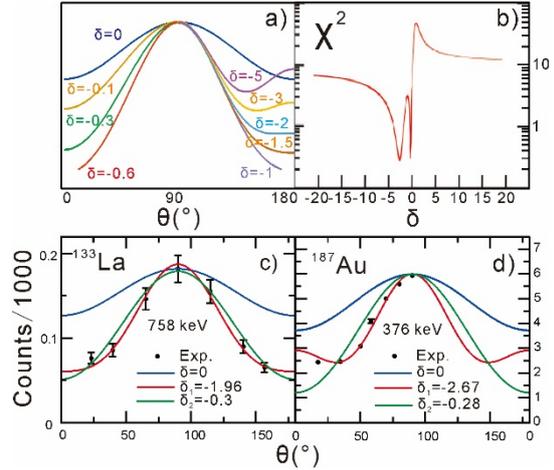

**Fig. 5 Two solutions on mixing ratio. a** Angular distribution curves with different $\delta$ values. Since all curves are symmetric about 90˚, only the left part for those with (-1<$\delta$<0), and only the right part for those with ($\delta$<-1) are shown to avoid overlapping. **b** A sketch figure of the $\chi^2$ as function of $\delta$. **c and d**, Estimated angular distribution curves and experimental results for the reported wobbling bands in $^{133}$La and $^{187}$Au. The curves for pure *M1* transition (in blue) and large $\delta$ (in red) are taken from the original figures in Refs. [13, 14], and those with smaller $\delta$ values and same $\sigma/I$ are also plotted (in green) for comparison.

Mixing ratios for the linking transitions of the low-spin wobbling bands in $^{135}$Pr[9,10], $^{133}$La[13] and $^{187}$Au[14] were extracted from angular distribution measurements, but the possible $|\delta_2|<1$ solutions were not mentioned or discussed, and the curves for $|\delta_1|>1$ were only compared with those for pure $M1$ transitions ($\delta=0$). Generally, the two solutions are expected to correspond to different polarization values, and therefore linear polarization ($P$) measurements can help to identify the correct one. Such measurements were performed for $^{135}$Pr[9] and $^{133}$La[13], but not for $^{187}$Au[14]. In addition, the $|\delta_1|>1$ solutions for two transitions in $^{135}$Pr were only selected based on their positive sign, but not on the magnitude of the measured $P$[9], which is insufficient since the $|\delta_2|<1$ solutions for those two transitions can also correspond to positive polarization values.

Alternative to angular distribution analysis, the $\delta$ values can be extracted using the measured directional correlation ratios of oriented states (DCO) and polarization values[33]. Such analysis was carried out for the transitions in $^{105}$Pd[11], $^{183}$Au[12], $^{133}$La[13], and $^{127}$Xe[15]. As shown in Fig. 5b, based on only measured DCO values two solutions for $\delta$ are possible, while the additional precise measured polarization can exclude one of them. For the polarization measurement, only the γ rays detected by two adjacent crystals of clover detectors are used, and consequently the uncertainty of the measured polarization value is relatively large. It is thus essential to carefully evaluate the uncertainty on the polarization value to confidently select the correct solution for $\delta$. However, in some published work, the uncertainties seem to be significantly underestimated. For example, a small uncertainty was reported for the polarization value of the 495-keV transition in $^{183}$Au[12], while the experimental data suggest much larger statistical errors (see Fig. 6 and Methods section), leading to larger uncertainty and insufficient precision to distinguish between the two possible solutions for $\delta$.

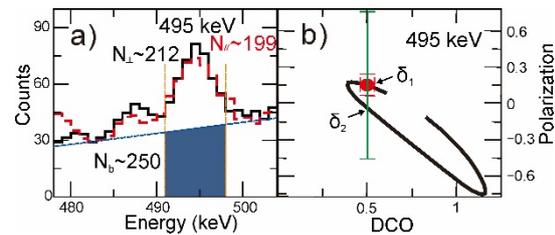

**Fig. 6 Statistical error on polarization for a reported transition in $^{183}$Au. a** Reported spectra used in polarization measurement for $^{183}$Au in Ref. [12]. $N_\perp$ and $N_{/\!/}$ represent the peak area of the 495-keV transition in spectra showing γ rays observed to scatter perpendicular or parallel to the reaction plane, respectively. $N_b$ is the background below the peak. The spectra are adopted from Fig. 1c in Ref. [11]. **b** Experimental (red symbol) and calculated (black line) DCO and linear polarization values for the 495-keV transition are taken from Fig. 2d in Ref. [11]. The error bar in green is the statistical error estimated from the spectra.

According to the overall assessment, the experimental results on several nuclei do not offer unambiguous evidence in support of the proposed low-spin wobbling bands.

The assigned wobbling nature was also based on the comparison between the experimental results on reduced transition probability ratios, (which strongly depend on the extracted $\delta$ values) and theoretical calculations. If the experimental $\delta$ values are considerably smaller than the theoretical expectation for wobbling bands, the wobbling interpretation is ruled out. On the other hand, only the measured $|\delta_1|>1$ values cannot constitute sufficient unambiguous evidence for the existence of wobbling motion. It is noticeable that in the work reporting the identification of low-spin wobbling bands, only interpretations in terms of signature partner band and wobbling band were considered. Actually, other mechanisms can also induce collective motion and thus induce enhanced *E2* component of the linking transitions, such as *β*-vibration, *γ*-vibration[34], and tilted precession[17,35]. Such collective excitations are well known in the low-spin domain.

**Discussion**

As an approximation, the wobbling motion is invalid at low spin. Therefore, its low-spin limit should be quantitatively investigated. On the other hand, low-spin precession is still expected to exist in nuclei, which need to be further investigated in a way other than wobbling. Actually, the interpretation in terms of tilted precession[17] of a low-spin rotational band was recently reported in $^{135}$Nd[35].

However, to make reliable conclusions on this topic, the problems hidden in the research paradigm should be solved. It is insufficient to exclusively adopt only the large $\delta$ solutions from the measured data and based on this to determine the type of excitation. In fact, the low-spin wobbling interpretation was adopted without exploring other possible modes of excitation. In further studies, the experimentally measured $\delta$ values, which are of critical importance for establishing the collectivity of the excitation, should be evaluated with better accuracy. The predicted measurable quantities, which can distinguish the different excitation modes, such as wobbling, tilted precession, vibrations, etc., need to be identified and properly treated in order to unambiguously assign the nature of such low-spin rotational bands.

**Methods**

*Details on the data analysis of $^{187}$Au*

In the present work, we extract the $\delta$ values of *M1/E2* transitions from the measured two-point angular-correlation ratio $R_{ac}$ and linear polarization $P$.

The $R_{ac}$ ratios were extracted from two *γ-γ* coincidence matrices which were built on prompt events, with 154˚ and 90˚ versus all angles. By setting the same energy gates in both matrices on the axis (all angles) and projecting on the other axis, gated spectra at 154˚ and 90˚ are created. The $R_{ac} = I_\gamma(154˚)/I_\gamma(90˚)$ ratios were calculated from the intensities of the *γ* rays of interest ($I_\gamma$) normalized by the respective efficiencies of the detectors at these two angles. Both the polarization and the angular correlation results are affected by the alignment of nuclei, described by $\sigma/I$. The $R_{ac}$ ratios depend on the multipolarity and on $\sigma/I$ of the initial state of the analyzed transition, but not of the gating transition. Better statistics can be achieved by summing several suitable gated spectra.

The polarization asymmetry ($A$) of Compton-scattered photons was measured following the prescription described in Ref. [33]. The asymmetry is defined as $A = (aN_\perp - N_\parallel)/(aN_\perp + N_\parallel)$, where $N_\perp$ and $N_\parallel$ denote the numbers of coincidence counts corresponding to photons scattered between the crystals of the clover detectors in the direction perpendicular and parallel to the emission plane, respectively. The quantity $a$ denotes the correction due to the geometrical asymmetry in the response of the clover segments, which was calibrated using standard gamma sources of $^{152}$Eu, $^{133}$Ba, and $^{60}$Co, and its dependence on energy is $a = 0.975(24) + 7.6(379) \times 10^{-6} E_\gamma$ ($E_\gamma$ is in keV).

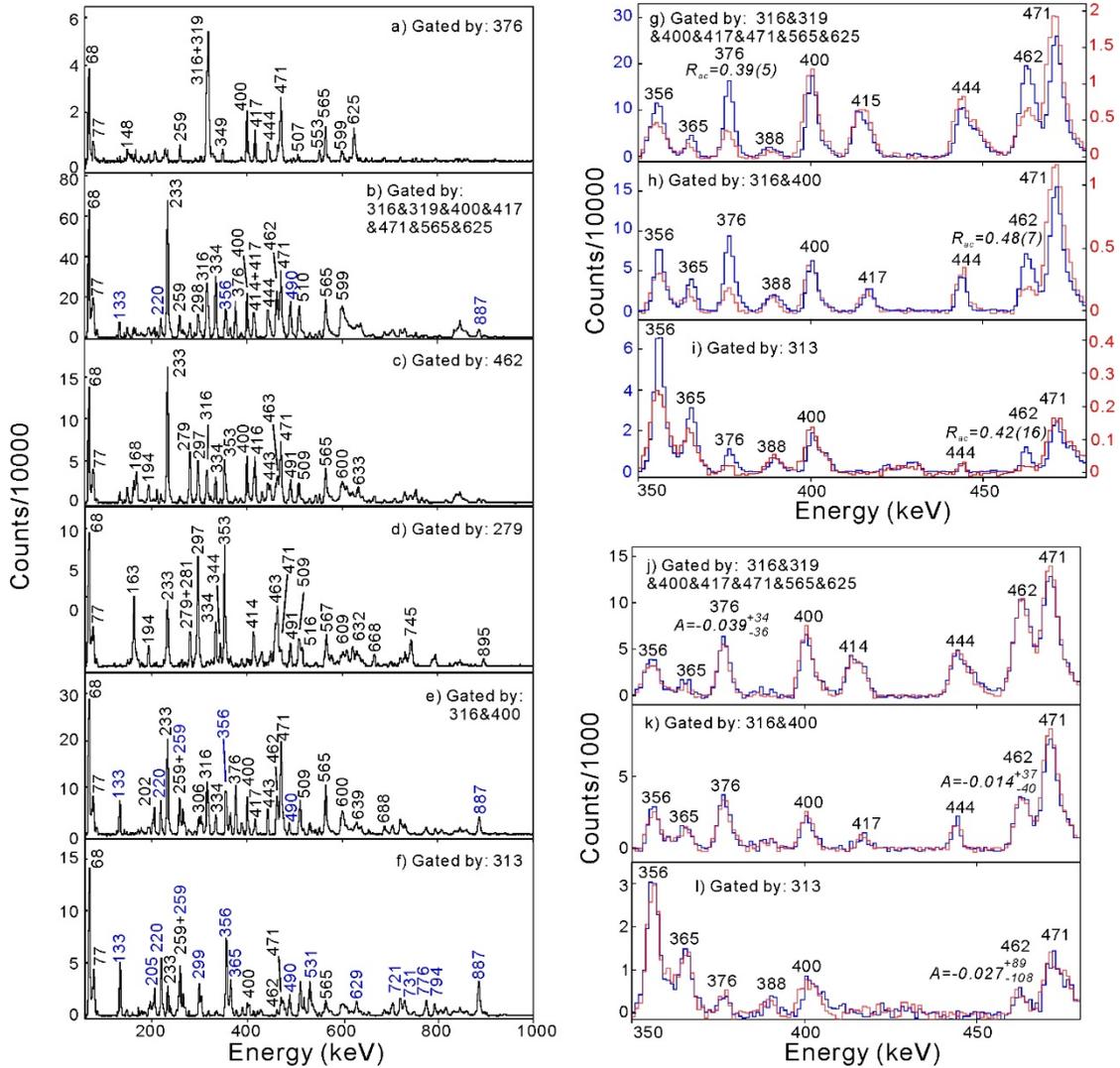

**Fig. 7 Typical spectra in the analysis. a-f**, Spectra produced from the symmetric matrix, where the contaminations from $^{188}$Au are marked in blue. **g-i**, Spectra used to deduce the $R_{ac}$ ratios for the 376- (**g**) and 462-keV (**h, i**) transitions. Spectra showing γ rays measured in the detectors at 154° and 90° are plotted in red and blue, respectively. **j-l**, Spectra used to deduce the asymmetries for the 376- (**a**) and 462-keV (**b, c**) transitions. Spectra showing γ rays observed to scatter perpendicular or parallel to the reaction plane are plotted in red and blue, respectively.

It is of crucial importance to select suitable gates to get reliable results. In order to deduce the $A$ and $R_{ac}$ ratios for the 376- and 462- keV transitions, we first checked the spectra gated by them in a symmetric $\gamma$-$\gamma$ matrix (see Figs. 7a and 7c). No contaminations were found in the spectrum gated by the 376-keV transition. We then selected seven strongest transitions in coincidence with it (the 316-, 319-, 400-, 417-, 471-, 565-, and 625-keV lines) to create the summed spectrum shown in Fig. 7b. The 462-keV transition instead can be contaminated by the close-lying 463-keV transition above the $T_{1/2} = 100$ ns isomer in $^{187}$Au (see Fig. 2 in the main text), and has to be carefully analyzed. The contamination from the 463-keV transition can be examined in Fig. 7d, where we show the spectrum gated by the 279-keV transition, which is located in the cascade above the isomer. Most of the strong peaks in the spectrum gated by the 462-keV transition also exist in the spectrum gated by the 279-keV transition, and therefore we have to avoid to gate on these transitions to deduce the $\delta$ values for the 462-keV transition. Comparing the two spectra gated by the 462-keV and 279-keV transitions, we can see that the 316- and 400-keV transitions appear only in the spectrum gated by the 462-keV transition. We therefore gated only on these two transitions to obtain the spectrum shown in Fig.7e. To eliminate the contamination from the 465-keV transition below the 400-keV transition in band 3, only the spectrum gated by the left part of the 316-keV peak (313 keV) was used. This spectrum is clean but has lower statistics (see Fig. 7f). Considering that the 465-keV transition has an $E1$ character, both the $R_{ac}$ ratio and the polarization asymmetry $A$ of the 462-keV transition can be overestimated by the contribution from the 465-keV transition. Therefore, the measured $R_{ac}$ and $A$ in the spectrum gated by the 316- and 400-keV transitions can be considered as upper limits for the $R_{ac}$ and $A$ values of the 462-keV transition. The peaks in blue in Figs. 7b, 7e, and 7f result from the coincidence with the 315-keV transition in $^{188}$Au, and do not affect the results for the 376- and 462-keV transitions.

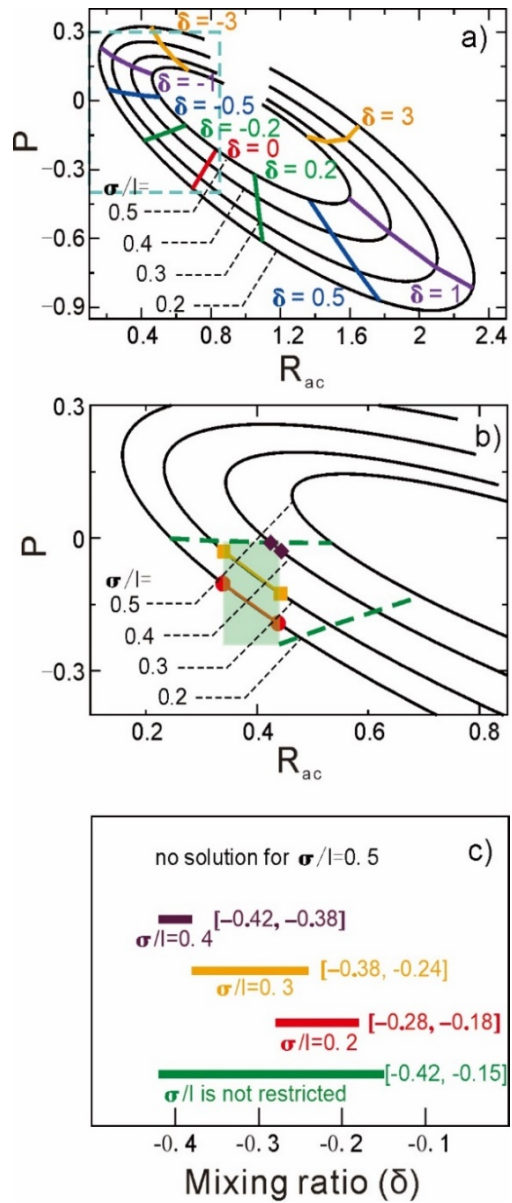

**Fig. 8 The method used to deduce the mixing ratio ($\delta$). a** The ellipse-like curves calculated as a function of $\delta$ with different $\sigma/I$ values, and the line-like curves calculated as a function of $\sigma/I$ with different $\delta$ values. **b** Different $\delta$ solutions for different $\sigma/I$ values, shown in black, yellow and red. The green rectangle represents the experimental uncertainties of $R_{ac}$ and $P$. The area within the dashed lines indicated in panel **a** is enlarged to show the details. **c** Comparison of the solutions for $\delta$ corresponding to different fixed $\sigma/I$ values, and the adopted range of $\delta$ corresponding to unrestricted $\sigma/I$ values.

The spectra produced by gating on selected transitions which are used to deduce the $R_{ac}$ ratios and polarization asymmetries $A$ are shown in Fig. 7. For the 462-keV transition, the upper limits of the $R_{ac}$ and $A$ values obtained from the spectra gated on the 316- and 400-keV transitions are adopted, in combination with the central values and lower limits deduced from the spectra obtained by gating on the left part of the 316-keV peak (313-keV).

The number of counts in the peaks is obtained by fitting background subtracted spectra. Partial uncertainties are induced by the selection of parameters for gating and fitting the peaks. To minimize such uncertainties, we used the same parameters for all spectra from which $R_{ac}$ and $A$ values are deduced in one fitting cycle. For each transition, we used the average $R_{ac}$ and $A$ values of those from at least 5 fitting cycles as their central values, and included the deviation into the total uncertainties.

The polarization $P$ is related to the polarization asymmetry $A$ through the formula $P = A/Q$, where $Q$ is the polarization sensitivity of the clover detectors. It is difficult to deduce $Q$ from the present work, since the strongest transitions are mainly populated from isomers. Therefore, we adopted the polarization sensitivity curve reported in Ref. [36], which was obtained for the same type of detectors as those used in the present work.

With the extracted $R_{ac}$ and $P$ values, the $\delta$ values are deduced as shown in Fig. 8. Actually, $R_{ac}$ and $P$ are functions of both $\delta$ and $\sigma/I$. The $\sigma/I$ values for the two transitions of 376 and 462 keV cannot be fixed from our measurement due to the existence of high-spin isomeric states. Therefore, the uncertainties on the $\delta$ values are based on the uncertainties of $R_{ac}$ and $P$ (indicated with the green rectangle in Fig. 8 b) and independently of the parameter $\sigma/I$, by considering all ellipse-like curves passing thought the green rectangle of the experimental uncertainties. For instance, the range of the extracted mixing ratio for the ellipse-like curves for $\sigma/I$ = 0.2 and 0.3 are shown in red and yellow in Fig. 8 b) and 8 c). Thus, the uncertainty range of $\delta$ (the green dashed lines in Fig. 8b) were derived for all possible values of the $\sigma/I$ parameter.

**Table 1 The single-particle orbitals**

| orbital label | sub-shell origin, $\Omega_{long}$ |
|---|---|
| #17 | ($h_{11/2}$; 7/2) |
| #18 | ($h_{11/2}$; 9/2) |
| #19 | ($h_{9/2}$; 1/2) |
| #20 | ($h_{11/2}$; 11/2) |
| #21 | ($h_{9/2}$; 3/2), ($f_{7/2}$; 1/2) |
| #22 | ($f_{7/2}$; 1/2), ($h_{9/2}$; 3/2), ($h_{9/2}$; 1/2) |
| #23 | ($h_{9/2}$; 5/2) |
| #24 | ($f_{7/2}$; 3/2) |
| #25 | ($h_{9/2}$; 7/2) |

*QTR calculations*

Quasiparticle-plus-triaxial-rotor model (QTR) couples an odd quasiparticle to a rotating triaxial core. The rotation is three-dimensional (3D), including rotations along all three major nuclear axes. Such

3D rotation can produce excited bands in odd-mass triaxial nuclei which result from a competition between single-particle excitations and collective motion, including precession.

We performed QTR calculations assuming γ = 12°, a value predicted for the $^{186}$Pt core by the theoretical calculations of Ref. [32]. Our calculations include the 9 negative-parity orbitals that are closest to the Fermi surface, as listed in Table 1. The Fermi level was found near the orbitals #19 and #20, as in Ref. [21], with dominant $h_{9/2}$ and $h_{11/2}$ nature respectively. Orbital #19 corresponds to the lowest-energy $h_{9/2}$ orbital, where the proton angular momentum is aligned along the short nuclear axis, and thus its projection along the long nuclear axis is $\Omega_{long}$ = 1/2. Orbital #20 corresponds to the highest-energy $h_{11/2}$ orbital with projection of the angular momentum $\Omega_{long}$ = 11/2. The orbitals #19, #21, #22 and #23 contribute to the wave functions of the states of the two $h_{9/2}$ bands (Bands 1 and 2).

The wave functions for five lowest-energy states of Band 1 (Band 2) in $^{187}$Au are illustrated in the top (bottom) panel of Fig. 9, where the wave functions are projected along the long nuclear axis. The largest contribution to the wave function of each state of Band 1, shown by the tallest blue columns in Fig. 9 (top panel), is attributed to the orbital #19 with $K_{long}$ = 1/2. This suggests that the corresponding rotation is orthogonal to the long axis, as $K_{long} = \Omega_{long}$. There is a small contribution from the orbital #21, (the second lowest-energy, $h_{9/2}$ orbital with $\Omega_{long}$ = 3/2) with $K_{long}$ = 3/2 (the red column in the top panel of Fig. 9).

Band 2 has dominant contribution from orbital #21 with $K_{long}$ = 3/2, corresponding to the tallest red column for each state in the bottom panel of Fig. 9. This suggests that for this band the odd proton occupies a different single-particle configuration associated with dominant contribution from the second lowest $h_{9/2}$ orbital, see Table 2. The rotation is again orthogonal to the long axis (as $K_{long} = \Omega_{long}$). The wave functions of the excited band also include a single-particle component similar to that of Band 1 (the blue column corresponding to the orbital #19 with $K_{long}$ = 1/2). At higher spins the contribution from the orbital #23 with $K_{long}$ = 5/2, indicated by the gray column, becomes more prominent. It shows a gradual misalignment of the odd proton away from the short axis because the orbital #23 has a dominant $\Omega_{long}$ = 5/2, $h_{9/2}$ nature.

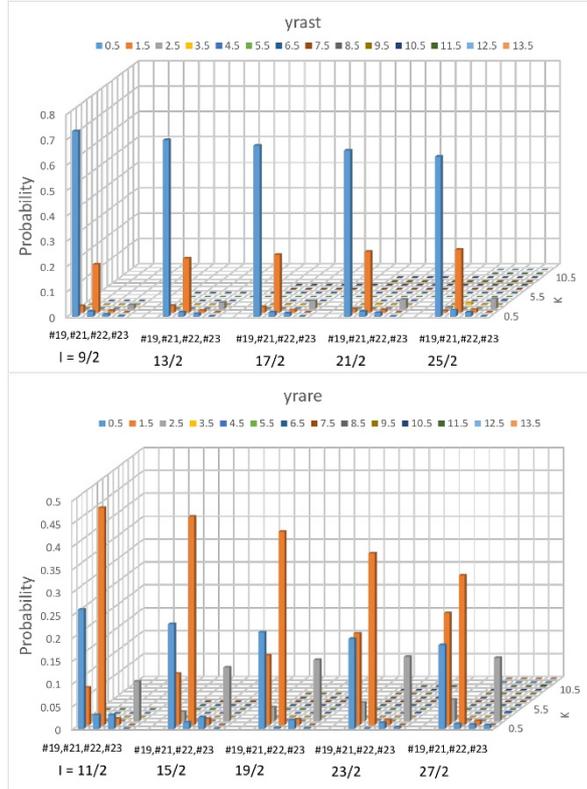

**Fig. 9 Contributions to the wave functions of five states of the yrast (top panel) and yrare (bottom panel) $h_{9/2}$ bands in $^{187}$Au.** The colour of the columns indicated in the legend shows the K projection on the long nuclear axis. Each state has single-particle contributions from orbitals #19, #21, #22 and #23, for more details see Table 2.

Another contribution, the orbital #19 and $K_{long} = 3/2$ shown by the second-tallest red column for each state, increases at higher spins. This contribution may describe nuclear precession as it corresponds to collective rotation not entirely along the intermediate axis because $K_{long} > \Omega_{long}$. In the calculated wave functions the contribution of the nuclear precession is small at low spins.

*Questioning the reported polarization results in $^{183}Au$*

The reported $\delta$ values in Ref. [12] are obtained from the analysis of the $R_{DCO}$ and polarization $P$ values[33], with the latter extracted using the formula $P = A/Q = (aN_\perp - N_\parallel)/(aN_\perp + N_\parallel)/Q$. However, no information on the two necessary calibrations, polarization sensitivity $Q$ and asymmetry correction $a$, has been reported.

We discuss mainly the statistical error on $P$ for the 495-keV transition using the reported spectra[12] (also shown in Fig. 6 of the present work) gated by 502-keV transition. It is the only gating transition which allows to avoid the contamination from the adjacent stronger 498-keV doublet peak. Considerable background is present below the 495-keV peak in both the perpendicular and the parallel spectra.

When calculating the uncertainties, one should always consider the errors induced by subtracting the background spectra obtained by gating on the Compton continuum near the gating transition. There is no information about these spectra, but as a conservative estimate for such weak peaks, one can assume that the subtracted background counts are not less than the background $N_b$ shown in Fig.6. Assuming $a=1$ and $Q_{495} = 0.233$ (adopting the calibration in Ref. [33]), the calculated statistical error of $P$ is larger than 0.61, estimated according to standard error propagation formulas[38]. This statistical error is much larger than the reported error ~ 0.09, suggesting a considerable underestimation. With an error of ~ 0.61 it is impossible to distinguish between the two solutions for $\delta$ (see Fig. 6), and therefore to determine the character of the 495-keV transition.

In addition, the proposed wobbling nature of the $h_{9/2}$ band in $^{183}Au$ was supported by the deduced $\delta$ value of another 498-keV connecting transition. However, the contamination from an adjacent 498-keV *E2* transition cannot be excluded by gating on any known transition in the published level scheme. Even if one wants to eliminate the component from the *E2* transition, the ratio of the intensities of the two transitions cannot be deduced directly. An estimated value of the intensity ratio would introduce a very large uncertainty on the deduced value of $P$. Therefore, the experimental results on the mixing ratios in $^{183}Au$ are not convincing, and do not represent a solid ground for the proposed wobbling interpretation.

**Data Availability**

The data that support the findings of this study are available from the corresponding author on reasonable request.

**Acknowledgements** The authors would like to thank the accelerator crew of HIRFL for providing the stable $^{18}$O beam during the pandemic of COVID-19. S.G. expresses his gratitude to Prof. P. W. Zhao for useful discussions. This work has been supported by the Strategic Priority Research Program of Chinese Academy of Sciences, (Grant No. XDB34000000), the National Key R&D Program of China (Contract Nos. 2018YFA0404402 and 2018YFA0404400), the Key Research Program of the Chinese Academy of Sciences (Grant No. XDPB09-02), the National Natural Science Foundation of China (Grant Nos. U1932137, U1732139, U1832134, 11775274, 11975209, and 11805289), the Cai YuanPei 2018 Project No. 41458XH, and the National Research Foundation of South Africa, GUN: 109134 and 116666.

**Author Contributions** S.G, X.H.Z prepared the proposal for the experiment, Y.D.F, H.Y.W, G.S.L, Y.H.Q and J.G.W set up the instrumentation, Y.Y.Y, H.J.O, J.B.M, S.W.X and Z. B tuned the beam through a secondary beamline, S.G, Y.D.F, G.S.L, Y.H.Q, J.G.W, Y.Z, B.D, W.Q.Z, A.R, K.R.M, H.L.F, J.F.H, J.H.L, J.H.X, B.F.L, W.H and Z.G.G monitored the detector, data acquisition and secondary beam systems, S.G, Y.H.Q, B.D, H.L.F, J.F.H, J.H.L and J.H.X carried out the data analysis and interpretation of the data, E.A.L, S.M, H.L.W, H.Y.M and M.L.L performed theoretical calculations, and S.G, X.H.Z, C.M.P, E.A.L, Y.D.F, M.L.L and Y.H.Z prepared the manuscript.

**Competing interests:** The authors declare no competing interests.